\documentclass[12pt,preprint]{aastex}
\received{Version 3 July 2009; revised 12 August 2009}
\accepted{15 September 2009}
\slugcomment{AJ, in press}
\shorttitle{Radio Supernovae in Arp 299}
\shortauthors{Ulvestad}
\def\H2{\ion{H}{2}}

\begin{document}

\title{Radio Emission from Young Supernovae and Supernova Remnants 
in Arp 299}

\author{James S.~Ulvestad}
\affil{National Radio Astronomy Observatory}
\affil{P.O. Box O, Socorro, NM 87801}
\email{julvesta@nrao.edu}

\begin{abstract}

We have made sensitive milliarcsecond-resolution radio images of 
the nearby merger galaxy Arp~299 at four epochs spread over 18 
months between 2003 and 2005.  The combined data revealed a total of 
30 point sources in the two primary merger nuclei. Twenty-five of 
these are found in the northeastern
nucleus (component ``A''=IC~694) over a region $\sim 100$~pc in
diameter, while 5 are in the southwestern nucleus (component 
``B1''=NGC~3690) within a region $\sim 30$~pc in size.  These objects 
are interpreted as young supernovae and supernova remnants; the ratio 
of the source counts in nuclei A and B1 is approximately equal to 
the ratio of their predicted supernova rates.  An approximate
luminosity function has been derived for nucleus A, and indicates that
it might contain as many as 500--1000 compact radio sources more
powerful than Cas~A; the integrated flux density of these sources would
be about 20\% of the total flux density seen at lower resolution.
A new supernova occurred in nucleus B1 in the first half
of 2005, having a peak radio power
at least 2,000 times the present power of Cassiopeia A.  This supernova
is located within 0.4~pc (projected distance) of an apparently older
supernova remnant, making it very likely that this indicates the presence
of a massive super star cluster within nucleus B1.  
Comparison of the typical radio flux densities of our compact radio
sources to the observed X-ray luminosities of nuclei A and B1
indicates that it is possible that one radio source in each nucleus
actually could be associated with an active galactic nucleus rather
than being a supernova remnant.

\end{abstract}


\section{Introduction}
\label{sec:intro}

It now appears likely that the most massive galaxies in the
universe formed from multiple mergers of smaller galaxies;
those mergers are apparent in the disturbed morphologies of
many galaxies imaged in deep optical and infrared fields such 
as the Hubble Deep Fields \citep{fer00}, the GOODS survey
\citep{dic03,gia04}, and the COSMOS survey \citep{sco07}.  Nearby and 
present-day analogs of these early galaxies often are found in 
Luminous and Ultraluminous Infrared Galaxies, many of which
are undergoing major mergers at the present epoch.  The nearest
major merger galaxy is generally regarded as the
`Antennae,' NGC 4038/9 \citep{whi99}, at a distance of 20~Mpc.  
Other nearby infrared-bright galaxies undergoing major mergers
are Arp~299 \citep{geh83,meu95,alo00}, 
NGC~3256 \citep{nor95,kot96,lip00}, 
and Arp~220 \citep{soi99,rov03,lon06,par07}.
The starbursts in these mergers and other galaxies obey the
so-called ``radio-infrared relation'', whereby their radio
and far-infrared powers are well-correlated
\citep{con92}; the infrared emission is thought to be
powered ultimately by the photons from hot young stars,
while the radio emission typically is powered by
synchrotron radiation from particles accelerated in
supernovae (SNe) and supernova remnants (SNRs).

In recent years, compact radio sources have been used as
one window providing insights into the nature of
merger/starburst galaxies.  In the most nearby starbursts,
such as M82 \citep{mux94,fen08} and NGC~253 \citep{ulv97,len06},
interferometers reveal a mix of thermal and nonthermal radio
sources representing \H2\ regions and fairly young 
SNRs.  However, in the somewhat more
distant merger galaxies, typically at distances of a few
tens of megaparsecs, use of the Very Long Baseline 
Interferometry (VLBI) technique is required to separate 
individual compact sources; the interferometer sensitivity 
on continent-scale baselines is insufficient to detect
\H2\ regions, so only non-thermal emission from young SNe
and SNRs is detected.  For example, a series of observations of the
merger galaxy Arp~220 \citep{rov03,lon06,par07} has detected
at least 49 objects in the two merger nuclei, provided the
radio spectra of many of these objects, and detected the 
explosion of several new supernovae.  

The subject of our investigation is the relatively nearby (41~Mpc)
merger galaxy Arp~299 (Mrk~171), which was noted as a starburst 
galaxy over 25 
years ago by \citet{geh83}.  The two primary nuclei of the merger
are the northeastern component A (often referred to as IC~694)
and the southwestern component B (often referred to as NGC~3690).
Although we focus primarily on the starburst in this paper, 
one or both nuclei also have been inferred to contain active 
galaxy components based on X-ray imaging and spectroscopy 
\citep{del02,zez03,bal04}, optical integral-field
spectroscopy \citep{gar06}, and the presence of nuclear H$_2$O
masers \citep{hen05,tar07}.

There is substantial CO emission associated with the Arp~299
starburst, and a relatively high ratio of HCN/CO at the two primary
nuclei, implying the presence of dense and warm molecular 
clouds in the two nuclei \citep{aal97}.  \citet{baa90} found
OH megamaser emission from nucleus A, and inferred that
this emission comes from a clumpy molecular medium located
100--600~pc in front of the continuum radio source(s) being
amplified.  Near-infrared imaging and modeling of Arp~299
by \citet{alo00} have shown that nucleus A accounts for
50\% of the near-infrared emission of the merger and has
a star-formation and supernova rate approximately six times
those of nucleus B.  The latter nucleus actually is 
resolved into a dominant component B1, and a weaker component
B2 located an arcsecond to the NW \citep{lai99,alo00,nef04}.
In order to provide context to the reader, we reproduce
two previously published figures here.
Figure~1 is an 814~nm {\it Hubble Space Telescope}
image of Arp~299, while Figure~2 is
a high-resolution Very Large Array (VLA) image of the
merger at radio wavelengths, both reproduced from \citet{nef04}.  

Arp~299 has hosted at least five optical supernovae in the last 
20 years, the last of which was the type II supernova SN 2005U 
\citep{mat05,mod05}.  These optical supernovae all 
occurred outside the primary merger nuclei; those
nuclei are likely to be the most frequent sources of 
supernovae in Arp~299, but their extreme obscuration makes
finding optical supernovae highly unlikely.  Even a two-epoch NICMOS 
near-infrared search failed to reveal any new supernovae
in the two merger nuclei \citep{cre07}.  However, high-resolution
VLBI imaging of Arp~299 revealed five compact radio sources in nucleus A, 
including at least one very young supernova \citep{nef04}.

This paper reports VLBI observations of
Arp~299 with considerably higher sensitivity than those
of \citet{nef04}, aimed at probing deeper into the radio 
luminosity function and searching for the onset of radio 
emission from young supernovae.  

\section{VLBI Observations}
\label{sec:obs}

The merger system Arp~299 was observed on four occasions using
an interferometer including the Very Long Baseline Array (VLBA) 
augmented by the Robert C.~Byrd Green 
Bank Telescope (GBT).  The VLBA \citep{nap93} is an interferometer 
consisting of 10 identical 25m antennas distributed from Hawaii
to the US Virgin Islands, while the GBT is a
100m x 110m single aperture in West Virginia, which was added to
the VLBA in order to approximately double the overall sensitivity
of the array.  Observations were carried out under program code
BU027, and took place at approximately six-month intervals between 
late 2003 and mid-2005.

Each of the four observing epochs consisted of 
almost identical 10-hr sessions.  At each epoch,
blocks of 60--110 minutes were alternated between sky
frequencies of 2.2715~GHz (hereafter 2.3~GHz) and 8.4215~GHz 
(hereafter 8.4~GHz).  There were four blocks at 2.3~GHz and
three blocks at 8.4~GHz; 2.3-GHz observing blocks were placed at 
the beginning and end of each 10-hr session in order to minimize 
the impact of atmospheric water vapor fluctuations in
low-elevation observing at 8.4~GHz.
At each band, four 8-MHz frequency channels
were used at both right and left circular polarizations.  All
were sampled at the Nyquist rate with two bits per sample, providing
a total data rate of 256~Mbit~s$^{-1}$.  These data were recorded on
a mix of instrumentation tapes and hard-disk modules 
which were shipped to Socorro, New Mexico for processing on the VLBA
correlator.  Only the parallel polarization hands were processed, in order to
keep the data rate within the correlator output limits; each 8-MHz 
frequency band was further subdivided into 32 or 64 spectral channels 
in order to preserve the capability for wide-field imaging.

The GBT was typically missing for 10\%--20\% of the time
in each observing session, due to software and pointing problems as
well as 5--10 minute periods to change between
2.3 and 8.4 GHz.  The Mauna Kea VLBA antenna did not observe in
the first and third epochs due to bad weather.  The Kitt Peak
VLBA antenna had 25\% of its 2.3 GHz data removed in all
observations, due to bad interference in one of the 
8-MHz bandpasses.  Additional data losses were typically
no more than 5\%--10\% of the data from 1--2 antennas for
a given observing run and frequency band.

The observations were carried out in
phase-referencing mode \citep{bea95}.  Observations of 
2.5~minutes in length on Arp~299 were interleaved with observations 
of approximately 45 seconds on the calibration source J1128+5925,
located 0.86$^\circ$ from Arp~299. Both the antennas and the
correlator were targeted midway between the strongest centimeter
radio component, nucleus A of Arp 299, and the possible background
source D, which were previously detected in the VLBI observations of
\citet{nef04}; this pointing preserved signal/noise in the
relatively small beam of the GBT as well as minimizing the
effects of bandwidth smearing.  Occasional 2.5-minute observations of 
J1127+5650, located 2.58$^\circ$ from J1128+5925, 
were made in order to provide a check of the 
phase-referencing quality.
Table~1 is a summary of the observations, including the observation
dates and the total integration time on Arp 299 at each frequency
band. 

All data calibration was carried out in NRAO's Astronomical Image 
Processing System, AIPS \citep{gre03}.  Standard corrections included
those for Faraday effects in the Earth's ionosphere, application of 
the most accurate Earth orientation parameters, gain corrections
to remove the effects of non-optimal sampler levels, and calibration of 
the feed rotation of the individual antennas.  Initial coarse solutions
for the interferometer delays were provided by fitting one or two 
scans of a strong calibrator (usually J0927+3902=4C~39.25), while 
final delay and phase calibration (or ``fringe fitting'')
was performed by least-squares fitting on each 
45-second scan of J1128+5925.  The resulting  delay and phase solutions
were interpolated in time to provide the final
calibration of Arp 299 and the check source.

Particular care was taken in setting the amplitude calibration
scale for the GBT. VLBA antennas provided useful system temperature 
data at roughly 1-minute intervals, and their calibration 
is judged to be accurate at the
5\% level.  However, the {\it a priori} and real-time calibrations 
of the GBT were less accurate; good values for GBT are important because it
is the most sensitive telescope in the observing array.  
Nominal values for GBT sensitivity were used to weight the GBT during 
the fringe-fitting process.  Then the amplitude scale was adjusted so that 
the flux densities of the amplitude-check sources J0854+2006 (OJ~287) 
and J1310+3220 varied smoothly with baseline length;  
in this adjustment, care was taken to make the amplitude on 
interferometer baselines to the GBT 
similar to those involving the relatively nearby VLBA
antenna in Hancock, New Hampshire.  This is a subjective
process because source structure is convolved with the fringe pattern
for each individual baseline, so the final
calibration was determined ``by eye'' using the source flux-density
plots.  Adjustments of the GBT sensitivity relative to the nominal
values were as high as 25\%, but more typically in the 10\% range.

\section{Imaging, Source Identification, and Flux-Density Measurements}
\label{sec:measure}

\subsection{Imaging}
\label{sec:imaging}

For each data set, images were produced using the standard deconvolution
and ``Clean'' algorithms in AIPS.  Initially, six fields were imaged in each data 
set, centered on the six strongest and most compact VLA sources detected
by \citet{nef04}.  These six regions were the two primary nuclei
of the galaxy merger, A and B1; source B2 to the NW of B1; the two possible 
additional merger nuclei, C and C$'$; and the possible background source D.
Each field consisted of a 1\farcs024-square image centered on the position
of the appropriate compact VLA source, with at least four pixels per 
resolution element.  The data sets were Cleaned to minimum
flux-density levels of 3--4 times the predicted noise for each image, in
order to assure that sidelobes due to imperfect sampling were removed
completely.

Careful source identification procedures, discussed in 
Section~\ref{sec:ids}, were carried out on all six fields.  The
only significant detections in individual epochs or in combined data
from all four epochs were found in the fields A, B1, and D.  Therefore,
to avoid generating spurious sources by Cleaning noise spikes, we
produced the final interferometer images by targeting only the three fields
with detected sources; this improved the noise characteristics of
the final images by $\sim 10$\%.  The noise levels from the final three-field 
images are cited in Table~1, and are typically 10\%--20\% above the 
theoretical values under ideal conditions.
The effective noise levels were somewhat higher
because of imperfect phase referencing, as discussed in 
Section~\ref{sec:flux}.  

After making the initial images, we experimented with self-calibration
in order to refine the atmospheric calibration and hence improve
the final images of Arp 299.  However, because there is very little 
correlated signal present, adequate signal/noise required averaging
times considerably longer than the atmospheric coherence
time, so self-calibration was not effective.  Thus, we depended
instead on the data with only the phase calibration relative to
J1128+5925.  

The purposes of the four epochs of observation were twofold--both to
achieve a time baseline long enough to see source flux density changes,
including the appearance of new sources,
and to achieve sufficient integration time to go significantly deeper
than the VLBA+GBT observations by \citet{nef04}.  For the latter
purpose, it is necessary to combine all the data from four epochs.
The traditional ways to do this are to (1) combine the calibrated data
from all four epochs in order to make a single set of images, or 
(2) average the four-epoch images in the image plane.  The former
method may be susceptible to errors in the case of significant source
variability, while the latter method does not easily retain the
relative weighting of the different baselines or of observations
with different noise and dynamic range levels.  We chose the first 
method, combining the calibrated data sets and then producing
three-field images of the combined data as for the individual epochs.
Figures~3 and 4 show the final 2.3~GHz images of the regions of 
nuclei A and B1 containing all the detected compact radio sources.
Images of the same fields at 8.4~GHz are not shown because of the 
smaller number of sources detected in that band.  From these figures,
we note that the compact radio sources in nucleus A extend over
a region $\sim 100$~pc in diameter, while the compact sources in
nucleus B1 are all found within a region $\sim 30$~pc in diameter.

\subsection{Source Identification}
\label{sec:ids}

The final source identification within the fields of A, B1, and D proceeded
in an iterative process.  First, all individual peaks above five times the
rms noise were identified in the 2.3-GHz images, and peaks above six times
the rms noise were identified at 8.4~GHz, both at the individual epochs
and in the combined data sets.  (The higher noise threshold at 8.4~GHz is
caused by the fact that the 8.4~GHz images have 16 times as many pixels
as the 2.3~GHz images.)
At some individual epochs, there were apparently
spurious sources above the cutoff flux densities near the strongest
($\sim 1$~mJy) sources, almost certainly due to the imperfect phase-referencing.
These spurious sources were easily identified because they occurred at 
different locations at the different epochs, and largely disappeared
when data sets were combined.

Locations of all sources above the cutoffs in the images from the combined 
data sets were inspected at
each individual epoch.  Sources exceeding the nominal cutoff in the
combined images were considered real if individual peaks above
$2.5\sigma$ also could be identified in at least two of the four epochs;
alternatively, if a peak above $7\sigma$ that was not an obvious
image defect could be found in a single
epoch, a source was considered real and possibly variable.  In fact, using these
criteria for the individual epochs, all sources above the cutoffs in 
the combined data sets were found to be real, and we judge the source
identification to be complete at the $5\sigma$ (2.3~GHz) and 
$6\sigma$ (8.4~GHz) levels for the combined data sets.

\subsection{Source Flux Densities and Positions}
\label{sec:flux}

For weak source detections, some sources can appear slightly resolved
because of noise plateaus lying near the actual sources.  In our
images, no sources were obviously resolved when this effect was
taken into account; given unresolved sources, the flux densities may
be measured by a variety of methods.  We chose to make quadratic fits to
the inner few pixels of each source in order to determine the peak
flux density, and used this value as the total flux density for the
apparently unresolved sources.  An alternative method of fitting 
beam-sized gaussians to each source makes assumptions about flux-density
distributions that may not be warranted, particularly in the
situation where imperfect phase-referencing scatters flux 
non-randomly in the image plane.

As stated previously, the phase-referencing process corrects imperfectly 
for the troposphere and ionosphere.  Thus, there is a net coherence loss 
in the Arp~299 images, and an apparent reduction in correlated flux 
densities.  We measured this reduction by using 
the phase-reference check source, J1127+5650 (hereafter J1127).  J1127
was imaged using the initial phase calibration applied from J1128+5925,
then self-calibrated and re-imaged.  The increase in peak flux density
by this procedure was a measure of the loss of coherence due to the
phase referencing errors.  The reduction in
flux density for Arp 299 at each epoch then was estimated by multiplying the 
coherence loss of J1127 by the ratio of separations from the
calibrator, ($0.86^\circ$/$2.58^\circ$), since the coherence
loss should depend approximately linearly on the separation between 
reference and target source \citep{bea95}.
Table~2 gives the flux-density correction factor for Arp~299
inferred by this method for each epoch, the average correction
factor over the four epochs, and the ``real'' noise levels achieved
for the combined data sets after the imperfect phase-referencing is 
taken into account.  Inspection of Table~2 indicates that
the 2.3~GHz coherence losses were more consistent from epoch to epoch,
while the 8.4~GHz coherence losses were much worse during the northern
summer; this supports the inference that the 2.3-GHz losses were 
dominated by an ionospheric calibration error whose magnitude is 
relatively independent of season, while the 8.4-GHz losses were
dominated by the troposphere, which is more variable 
during the summer.  Final flux densities for the
Arp~299 sources were obtained by multiplying the fitted flux
densities by the amplitude correction factor at each epoch, or by the
average correction factor for the combined data sets.

Errors in flux-density estimates have three different causes.
First, there is an overall scale error, dominated by the error
in calibrating the GBT, which we estimate to be 10\% (see 
Section~\ref{sec:obs}).  Second, there is an error
in estimating the coherence loss, including the assumptions
that this loss depends linearly on target/calibrator separation, is
otherwise independent of direction, and has the same temporal structure
for Arp 299 and J1127 (the latter source was observed only every hour
or two).  We take this error to be 50\% of the inferred coherence loss;
i.e., for an apparent coherence loss of 15\%, the $1\sigma$ error in
the loss factor is assumed to be 7.5\%. (In reality, the error
in coherence loss is surely non-gaussian, but there is no definitive
reason to adopt any other form.)
  Third, there is the error caused by the limited signal/noise
ratio of the data, which is as high as 20\% for a $5\sigma$ source and
as low as 3\% for a 1-mJy source at a single epoch.  We add these three
errors in quadrature to derive the final errors at each epoch. 
For the final images using the
combined data sets, we assumed that the flux-density scale errors and
coherence-loss errors were uncorrelated from epoch to epoch, implying that
they are reduced by a factor of two for four epochs. 

Source position measurements are unaffected by amplitude scale errors.
At 2.3~GHz, epoch-to-epoch position shifts should be dominated by 
signal/noise considerations, which limit the positional accuracy to 
approximately the beam size divided by the signal/noise ratio; this
may be as large as 1~mas for a $5\sigma$--$6\sigma$ source.  At 8.4~GHz, 
position shifts between epochs sometimes are larger than the 0.2--0.3~mas
that this simple assumption would indicate, most likely due to
the imperfect phase calibration.  We must also
account for the uncertainty in the position of J1128+5925; errors
are given by \citet{fey04} as 0.16~mas in right ascension and 0.27~mas
in declination.  In fact, the current VLBA calibrator 
list\footnote{({\em http://www.vlba.nrao.edu/astro/calib/vlbaCalib.txt\/})}
was used; it quotes 
50\% larger errors, and uses updated coordinates that differ from 
\citet{fey04} by about 0.05~mas in each coordinate.
Given that
the signal/noise at 2.3~GHz limits position determination to 1~mas accuracy 
for the weakest sources, and that the phase-referencing may limit
8.4~GHz positions at a similar level, we estimate position errors of
1.0~mas in each coordinate, and quote absolute source positions to an 
accuracy of 0.1 milliseconds of time in right ascension and 1.0~mas 
in declination.

Table~3 gives the final positions and flux densities of all our confirmed
sources at 2.3~GHz, while Table~4 gives a similar list for 8.4~GHz.
In these tables, source names are given by convention as position offsets from
$\alpha_{\rm J2000} = 11^h28^m$ and $\delta_{\rm J2000} = 58^\circ 33'$;
for sources in common between both frequency bands, the source name 
derived from the 8.4 GHz position is used.
The source properties and their implications are discussed further in
section~\ref{sec:sources}.

\section{Discussion of Compact Radio Sources}
\label{sec:sources}

\subsection{Source Detection Summary}
\label{sec:detections}

Inspection of Tables~3 and 4 shows that there are significant 
milliarcsecond radio sources detected in both of the primary
nuclei of the galaxy mergers, A and B1; as stated in 
Section~\ref{sec:imaging}, there were no detections in 
B2, C, or C$'$.  Source D is detected at all epochs, but with
no other radio detections nearby.  \citet{nef04} previously 
reported five VLBI detections in A, as well as detection of
D, but none in B1.  The much deeper observations described in
this paper revealed 19 sources at 2.3~GHz in A, as 
well as 13 sources at 8.4~GHz; since only 7 objects were detected
at both frequency bands, there are now a total of 25 distinct
VLBI radio sources in A.  In addition, there were four ``steady''
detections in B1, plus a variable source that only appeared
at the fourth epoch.
Since no sources are resolved at either 2.3 or 8.4~GHz,
we infer that they generally have sizes smaller than 
one milliarcsecond, or 0.2~pc.  The most likely interpretation,
by virtue of analogy with similar sources in other merger
or starburst galaxies such as Arp~220, M82, and NGC~253, 
is that the sources are all
young, compact SNe or SNRs.  In fact, the
ratio of 5:1 in source counts between A and B1 is very close
to the ratio of 5.9:1 in their estimated supernova rates
\citep{alo00}, indicating that the stellar mass functions
and supernova evolution may be similar between the two nuclei.
In an extreme merger such as Arp~299, where much of the
starburst has occurred in the last 10~Myr \citep{alo00},
it is likely that virtually all of the radio sources have
resulted from Type~II supernovae.

\subsection{Source Variability}
\label{sec:var}

In Figure~5, we plot the four-epoch 2.3~GHz data and 8.4~GHz
data for the strongest source at each band, one of intermediate 
strength, and one of the weaker sources.  This figure includes
data for only a few sources, but most of the
detected sources show no significant variability.
Since the error bars at individual epochs are
typically in the 10\%--15\% range, the upper limits on variability
over 18 months are only in the vicinity of 20\% for
the non-variable objects.  These limits are consistent with the much
smaller variability and limits seen in most radio sources in NGC~253,
M82, and Arp~220 \citep{ulv94,ulv97,kro00,par07,fen08}.

We find only two sources in our program that display significant 
variability, both at 8.4~GHz.  The first is the strongest 8.4~GHz 
source shown in the right-hand panel of Figure~5, designated as
33.621+46.71 in Table~4; this source 
in nucleus A appears to have a systematic flux density decline 
over four epochs.  It was initially reported by 
\citet{nef04} as a very young radio supernova, 
and found to have an 8.4~GHz flux density of 3.2~mJy 
at epoch 2003.11; this higher flux density in early 2003
confirms the monotonic decline seen in Figure~5.

The other variable source at 8.4~GHz appeared only at the fourth 
epoch, in nucleus B1, as noted in Table~4 (source 30.988+40.78).  
Further discussion
of this source is deferred until Section~\ref{sec:RSN}.

\subsection{Source Spectra}
\label{sec:spectra}

Figure~6 shows the distribution of two-point spectral indices 
($\alpha$, defined by $S_\nu\propto \nu^{+\alpha}$) for all
the detected sources except the new source that appeared at 
the fourth epoch; those sources with upper limits to their
flux densities at 2.3 or 8.4 GHz are shown hatched.
Errors in the values of $\alpha$ range from 0.09 to 0.19 for
the sources detected at both frequencies, while the errors
on the spectral-index limits for sources detected at only
one frequency range from 0.16 to 0.21.
Because only 8 of the 29 sources in this plot were detected
at both frequencies, we have not attempted to find
the true distribution of source spectra. Mathematically,
one could use survival analysis to derive a hypothetical ``true''
spectral index distribution.  However, this would rely on an
implicit assumption that sources with upper limits at each
frequency are distributed similarly to the sources detected
at both frequencies, and this is very unlikely to be a correct
assumption.

We can say that at least seven of the sources have positive spectral
indices, almost surely indicating optically thick synchrotron
emission.  (Brightness temperatures of these milliarcsecond
sources are above $10^6$~K for even the weakest sources, and thus 
the radio emission is highly unlikely to be thermal in nature.)
The two variable sources, only one of which is included in
Figure~6, have spectral indices
$\alpha > +1.5$ between 2.3 and 8.4~GHz.  Such steep positive
spectra are characteristic of very young radio SNe
which have become optically thin at the higher frequency,
but remain optically thick at the lower frequency \citep{wei02}.
Hence we infer that the radio sources occupying
the right side of Figure~6 are likely to be the most recent
supernovae.

\subsection{Radio Luminosity Functions}
\label{section:rlum}

The distributions of radio luminosities at 2.3 and 8.4~GHz
are shown in Figure~7, with the objects from nuclei A and B1
distinguished from one another.  The detection thresholds
are $1.7\times 10^{19}$~W~Hz$^{-1}$ at 2.3~GHz and 
$1.8\times 10^{19}$~W~Hz$^{-1}$ at 8.4~GHz.  For reference, 
we compare these values to the galactic SNR
Cassiopeia~A.  Using the spectrum and flux-density evolution of Cas~A
found by \citet{baa77}, as well as the 3.4~kpc distance derived
by \citet{ree95}, we derive Cas~A powers at epoch 2005.0 of 
$1.6\times 10^{18}$~W~Hz$^{-1}$ at 2.3~GHz and 
$6.1\times 10^{17}$~W~Hz$^{-1}$ at 8.4~GHz.  Thus, our
respective detection thresholds for milliarcsecond radio 
sources in Arp 299 are $\sim 11$ and $\sim 29$
times the Cas A power at 2.3 and 8.4 GHz.

It is of interest to compute the total number of radio 
sources that might be expected above the Cas~A power in
Arp~299, as this may provide some insights into the
time evolution of the sources.  We consider only nucleus A,
since the number of radio sources in nucleus B1 is too small
to derive useful statistical results.  One can use the
distribution of observed radio powers shown in Figure~7 
to fit a luminosity function, and then integrate that luminosity
function down to the power of Cas~A.  However, the small-number statistics
give rather large errors on any such fit.  For a rough estimate,
we assume a power-law form for the luminosity function, and
take bin widths of 0.2 dex in radio power.  At 
$\log(P) = 19.75$ (in W~Hz$^{-1}$), we ameliorate small-number
statistics slightly by averaging this bin with the two adjacent bins,
and find 2 sources per bin at 8.4~GHz and 2.33 sources per bin at 2.3~GHz.
A two-point fit between these average values at $\log(P)=19.75$ 
and the values at $\log(P) = 19.35$ gives the following results,
where $P_{19} = P/(10^{19}~\rm{W~Hz}^{-1}$):
\begin{equation}
dN(P_{19})/d\ln P_{19}\ \sim\ 80{P_{19}}^{-2.6}
\end{equation}
at 2.3 GHz, and
\begin{equation}
dN(P_{19})/d\ln P_{19}\ \sim\ 30{P_{19}}^{-2.2}
\end{equation}
at 8.4 GHz.
Integrating these estimated luminosity functions down to the 
luminosity of Cas~A at the respective frequencies, we find
that nucleus~A of Arp~299 would contain approximately 890 
sources stronger than Cas~A at 2.3~GHz and 740 sources stronger
than Cas~A at 8.4~GHz.  The error bars on these numbers
are roughly a factor of two, based on relatively small-number 
statistics, and also ignore the possibility that the shapes of
the luminosity functions might change at lower luminosities.
Indeed, it is possible
that most of the objects detected in our observations are
fairly young radio SNe, whose emission still is largely
dominated by interaction with their own mass-loss shells,
rather than SNRs dominated by interaction with an 
external medium \citep{che01,che04}; any transition from SNe
to SNRs most likely would cause the luminosity function to change
shape.  Extensive discussion of
the SNe/SNR evolution in the merger Arp~220 is given by 
\citet{par07}, and we refer the interested reader to that
paper for details that also may apply to Arp~299.

The summed flux densities of the observed young supernovae and
supernova remnants in nucleus A (Tables~\ref{tab:sflux}
and \ref{tab:xflux}) amount to 5.3~mJy at 2.3~GHz
and 4.2~mJy at 8.4~GHz.  Multiplying the above luminosity
functions by power and integrating down to the power of Cas~A,
we find respective total flux densities of 19 and 14~mJy at
2.3 and 8.4 GHz, in compact radio sources more powerful than 
Cas~A.  \citet{nef04} gave respective flux 
densities of 101 and 77~mJy at 4.9 and 8.4~GHz in the sub-arcsecond
component of nucleus A.  Thus, even if the estimated luminosity functions
continue unbroken to the power of Cas~A, the total flux density in 
young SNe and SNRs more powerful than Cas~A is only 
$\sim$15--20\% of the total radio flux
density in nucleus A, consistent with all observations.

We can consider the implications of this result by comparing
it with the expected supernova evolution in our own Galaxy. 
It has been $\sim 330$~yr since the supernova event that 
created Cas~A \citep{hug80}, so one can examine the (admittedly
naive) hypothesis that all supernovae in Arp~299 evolve exactly 
like Cas~A.  For a supernova rate of 0.65~yr$^{-1}$ in
nucleus A of Arp~299 \citep{alo00}, there should now be
$\sim 200$ SNRs younger than Cas~A in that
nucleus, while integration of the estimated 
luminosity functions implies that Arp299-A may contain
as many as $\sim 800$ SNRs above the Cas~A power.
This is weak evidence that the
SNRs in Arp 299 stay stronger for a longer period of
time, as one might expect if they explode in a dense
medium and expand more slowly than in our own Galaxy.
By comparison, the nearby starburst galaxies M82 and  NGC~253 
have $\sim 10$--20 radio SNRs
more powerful than Cas~A \citep{ulv97,fen08} in galaxies with 
supernova rates of $\sim 0.1$~yr$^{-1}$, which would scale to
$\sim 60$--120 objects for supernova rates of
$\sim 0.6$~yr$^{-1}$. This implies that the 
SNR evolution may be more rapid, or the SNRs intrinsically
less luminous, in these weaker starbursts.  Alternatively, 
as suggested for Arp~220 by \citet{par07}, it may be that the
initial mass function in the more extreme starburst galaxies is
more top-heavy.  However, the evidence for the existence of
such top-heavy mass functions is inconclusive \citep{elm05}.

Given the large uncertainties in the
luminosity function, a significantly deeper VLBI observation
of Arp~299, which should be possible within 2--3~yr, would be
required to test whether the young SNe and SNRs in Arp~299
are more radio-powerful or longer-lasting than those in
less active starbursts.  A turnover in the luminosity
function that would be caused by a top-heavy initial mass function,
as well as additional young supernovae, might
be revealed by such an observation.

\subsection{The Two Newest Supernovae}
\label{sec:RSN}

As mentioned in Section~\ref{sec:var}, we have detected two
compact variable sources at 8.4~GHz, which we identify as
young supernovae.  First, we consider the source 33.621+46.71,
previously denoted as A0 by \citet{nef04}.  This object
had an 8.4~GHz flux density of 3.2~mJy at epoch 2003.11,
and fell to 1.1~mJy at epoch 2005.55.
Fitting an exponential falloff of the form
$S(t)=S_0(t_0)\ (t/t_0)^{-b}$ since epoch 2003.11, we find
that the source flux density at 8.4~GHz was declining as 
$t^{-0.63}$.  However, since the source was not detected
at 2.3~GHz at any of our four epochs, we conclude that it
still remained optically thick at that frequency 2.5~yr 
after becoming optically thin at 8.4~GHz.  The slow rise
time at 2.3~GHz is consistent with a Type~II
supernova, but not a Type~Ib/c supernova (see Table~3
of Weiler et al. 2002).

The other variable source was detected only at our last epoch, 2005.55, 
in Nucleus B1; it had an 8.4-GHz flux density of $1.17\pm 0.28$~mJy
at that time.  We have no later VLBI observations to study its evolution,
but we did search for all relevant observations in the
VLA data archive.  We found two {\bf A} configuration
observations that were taken subsequent to the
many epochs of data analyzed by \citet{nef04}.
These observations were made under program code AC749, on 
2004NOV02 (2004.84) and 2006APR15 (2006.29), and have resolution
of 0\farcs2.  We have extracted both 
data sets from the VLA archive and analyzed them.
Historically, the 8.4~GHz flux densities of nucleus B1 ranged from 
5.9 to 7.2~mJy in {\bf A} configuration data obtained between
1990 and 2002, while nucleus A ranged from 70 to 75~mJy \citep{nef04}.
Thus, a new radio source with a flux density of a few millijansky or
less would not be noticeable within nucleus A, but might be
detectable in nucleus B1.
Table~5 gives the results of our flux-density measurements of
nuclei A and B1 from the archival VLA data.  We find that the
flux densities of both nuclei were consistent with their
historical values at epoch 2004.84, whereas the 8.4-GHz flux density
of B1 had approximately doubled in 2006.29.

Our best estimate for the change in the flux density of B1 is an 
increase of approximately 6.5~mJy between 2004.84 and 2006.29.  
Given the estimated supernova rate of only $\sim 0.11$~yr$^{-1}$ in 
B1 \citep{alo00}, it seems reasonable to attribute this entire 
increase to the single new source we have detected in our 
VLBI observations.  Given that assumption, Figure~8 shows an 
8.4~GHz light curve for the Arp~299 radio supernova, using the 
combination of archival VLA data and VLBI observations. 
The supernova in Arp~299 reached a peak power of at least 
$1.2\times 10^{21}$~W~Hz$^{-1}$ at 8.4~GHz, about 2,000 times the
Cas~A power.  This is comparable to the peak radio powers
of the luminous Type~IIn supernovae summarized by 
\citet{wei02}.  It is far more powerful than the late-time powers
of Type~Ibc supernovae \citep{sod06}, and nearly 10 times
more powerful than the recently discovered radio supernova
in M82 \citep{bru09}.

We note the estimated supernova rates of 0.65~yr$^{-1}$ for nucleus 
A and 0.11~yr$^{-1}$ for B1 \citep{alo00}.  Our 8.4-GHz VLBI observations 
and those of \citet{nef04} spanned a total of $\sim 2.5$~yr; detection of
one new supernova in each of A and B1 is consistent with those values,
assuming that most of the young supernovae actually are 
detectable radio supernovae.

\subsection{Supernovae as Signposts for Super Star Clusters}

The new supernova in B1 appeared within two milliarcseconds, or 0.4~pc 
(projected), of a steady 2.3~GHz source with a flux density of
0.25~mJy; this 2.3~GHz source did not vary when the new source
appeared.  The four steady 2.3~GHz sources all lie within a rectangle
about 100 by 160~mas, with a total area of 
$1.6\times 10^4$~mas$^2$.  One can test the hypothesis that a new
supernova occurring with equal probability at any location in this
rectangle would appear very close to an existing source by chance.
For a completely random location of the new supernova, the {\it a priori} 
probability that it would occur within 5~mas of an existing source
is 2.0\%, and the probability of occurrence within 2~mas of an
existing source is 0.3\%.  Thus the likelihood is that the new supernova
is physically associated in some way with the existing radio source.
If the two radio sources actually are SNe/SNRs
separated by 0.4~pc, their supernova shells would overlap in
no more than 100~yr even for expansion speeds of only 2,000~km~s$^{-1}$.

The most logical physical association is that both the older SNR and
the new supernova are located in the same super star cluster (SSC) in 
nucleus B1.  Such SSCs have typical radii of 4~pc in the Antennae 
\citep{whi99}, corresponding to 20~mas at Arp~299.  In fact, 
the SSCs found nearby, such as R136 in 30 Doradus
\citep{hun95}, are highly centrally condensed, so massive stars and
supernovae may be most likely to coexist over a much smaller region.
Since the compact VLBI sources detected in B1 are otherwise separated 
by typical distances of 10--15~pc, it is tempting to hypothesize
that each is a beacon identifying the location of an individual
SSC.  We note that a number of SSCs dominated by thermal emission 
previously have been detected and imaged by the VLA 
\citep{tur00,bec02,joh03,tur04}.  However, the SSCs dominated by 
thermal emission apparently are younger than 3~Myr,
whereas the possible SSCs indicated by our VLBI observations 
must be older than 3~Myr in order to harbor SNe and SNRs.

Statistical inferences based on single objects are, of course, quite 
perilous.  However, \citet{nef04} also noted that the new 8.4-GHz source in
nucleus~A appeared within 3~pc (projected) of a previously detected 
2.3-GHz source, and suggested that they might be in the same
SSC.  There are at least four compact radio sources 
in nucleus~A within a projected
separation of $\sim 10$~pc, making it likely that these are
associated in some way.  A deeper VLBI integration would reveal
whether there are weaker SNRs physically associated with the same
regions.

The possible confusion of radio
sources at different frequency bands provides some cause
for anxiety in considering the spectral indices of the eight objects 
apparently detected
at both radio frequencies (see Section~\ref{sec:spectra}), because
there is no guarantee that detections at multiple frequencies
actually correspond to the same source.  The eight multi-frequency
objects do not vary significantly at 8.4~GHz over the course of 
1.5~yr. Most appear to have optically thin synchrotron
spectra, and thus are likely to be older SNRs radiating at both 2.3 
and 8.4~GHz rather than chance coincidences.  However,
object 33.630+46.79 has a flat spectrum 
between 2.3 and 8.4~GHz, so it is possible that it is an SNR
with an unusual spectrum or that it does, in fact, correspond
to two confused sources.  

\subsection{Possible Active Galactic Nuclei}

In recent years, there have been several reports that one or both
of Arp 299 nuclei A and B (or B1) may be active galactic nuclei
(AGNs). \citet{del02} detected a hard X-ray component in Arp~299;
arcsecond-scale imaging by \citet{zez03} indicated that nucleus
B1 has a hard X-ray spectrum and may be an AGN, while spectral
fits to nucleus A indicated that it too could have an AGN component.
\citet{tar07} also found H$_2$O maser emission in both A and B1,
supporting the possibility of dual AGNs in Arp~299.

Since we detect numerous radio sources in each nucleus, it is not
possible to determine which (if any) might correspond to the X-ray
sources.  However, we can at least check whether the radio/X-ray
ratios for our typical compact radio sources might be consistent
with the presence of AGNs.  We make use of the definition 
$R_X = \nu L_\nu ({\rm 5~GHz})/L_X ({\rm 2-10~keV})$ 
\citep{ter03}. In nucleus B1, the X-ray source
has an X-ray luminosity of $\sim 7\times 10^{39}$~ergs~s$^{-1}$ in
the 0.1--10~keV range \citep{zez03}; our typical compact radio
source (excluding the single young supernova) 
has a strength of $\sim 200$~$\mu$Jy at 5~GHz (based on the 2.3~GHz
detections and spectral limits), giving 
$\nu L_\nu\sim 1\times 10^{36}$~ergs~s$^{-1}$, and 
$\log R_X \sim -3.8$ (ignoring relatively small corrections for
the X-ray spectrum).  The X-ray source in nucleus~A emits 
$\sim 1\times 10^{40}$~ergs~s$^{-1}$ \citep{zez03}; the relatively
strong 1~mJy radio source which \citet{nef04} speculated to be a 
possible AGN has 
$\nu L_\nu\sim 3\times 10^{36}$~ergs~s$^{-1}$ at 5~GHz, so 
$\log R_X \sim -3.5$ if this radio source were associated with
the X-ray emission, or 
somewhat lower (similar to Nucleus B1) if one of the weaker
radio sources were the X-ray source.  Thus our estimated values of
$\log R_X \sim -4.0$ to $-3.5$ are consistent with the values
of low-luminosity Seyfert galaxies observed by \citet{ter03},
and it is possible that one of the compact radio sources in
either A or B1 (or both) may be an AGN.  However, it also is 
possible that an AGN could have $\log R_X \sim -4.5$, and
hence be undetected in the present VLBI observations.
The best way to determine which of the compact radio sources might
be an AGN may be to conduct a high-sensitivity phase-referenced 
VLBI observation of the H$_2$O masers in order to see if they
could be identified with any of the compact continuum radio
emitters.

\citet{nef04} hypothesized that VLBI
resolution of source D could indicate whether it belongs to 
Arp~299 or is a background AGN.  None of our four epochs of 
observations have resolved this radio source, so we suspect
that the possible resolution on the scale of the beam size 
that was reported previously \citep{nef04} may have been spurious.
In fact, given the X-ray luminosity \citep{zez03} and average
radio flux density of this object, it has $\log R_X\sim -2.3$, 
consistent with values found for some PG quasars by \citet{ter03}.

\section{Summary}

We have observed the nearby galaxy merger Arp~299 at four epochs
and two frequencies with a long-baseline radio interferometer 
consisting of the VLBA and the GBT.  The primary results of
these observations are as follows:

\begin{enumerate}

\item Thirty compact radio sources were detected in the two
primary merger nuclei above limits of 10--30 times the 
luminosity of Cas~A. 
The 25 detections in nucleus A span a diameter of 
$\sim 100$~pc, while the 5 detections in nucleus B1
span a diameter of $\sim 30$~pc.

\item The ratios of the numbers of radio sources in nuclei A
and B1 are consistent with the ratios of the supernova rates
inferred by \citet{alo00}.

\item Most sources have variability upper limits of 
$\sim 20$\%  over 18.5 months, consistent with
the interpretation that they are relatively young supernova remnants.

\item A few objects were detected at 8.4~GHz and not at 2.3~GHz,
indicating that they may be younger radio supernovae that are still
optically thick at the lower frequency.

\item A previously detected young supernova in nucleus A had
a steadily declining 8.4~GHz flux density over 2.5~yr, but still
had not become detectable at 2.3~GHz after the same interval.

\item A new supernova was detected in nucleus B1, with an
apparent explosion date in the first half of 2005; this 
object had a peak luminosity at least 2,000 times the Cas~A power.
The new supernova occurred within 2~mas (0.4~pc projected)
of a previously known
steady radio source; it is highly likely that these two
radio sources are the signposts of a super star cluster that
formed at least 3~Myr ago in nucleus B1.

\item Comparison of the radio powers of the individual sources
to the X-ray luminosities of nuclei A and B1 indicates that
it is possible that one radio source in each nucleus is actually
an active galactic nucleus rather than a young SNR.

\end{enumerate}

\acknowledgments

The National Radio Astronomy Observatory is a facility of the 
National Science Foundation operated under cooperative agreement by Associated 
Universities, Inc.  I thank the many NRAO staff that made these 
observations possible.  I also thank Susan Neff for assistance
in early stages of this project, and for supplying figures from
\citet{nef04}. Finally, I thank the anonymous referee for suggestions
that significantly improved the presentation of the paper.

{\it Facilities:} \facility{VLBA}; \facility{GBT}; \facility{VLA}.

\clearpage

\begin{figure}
\plotone{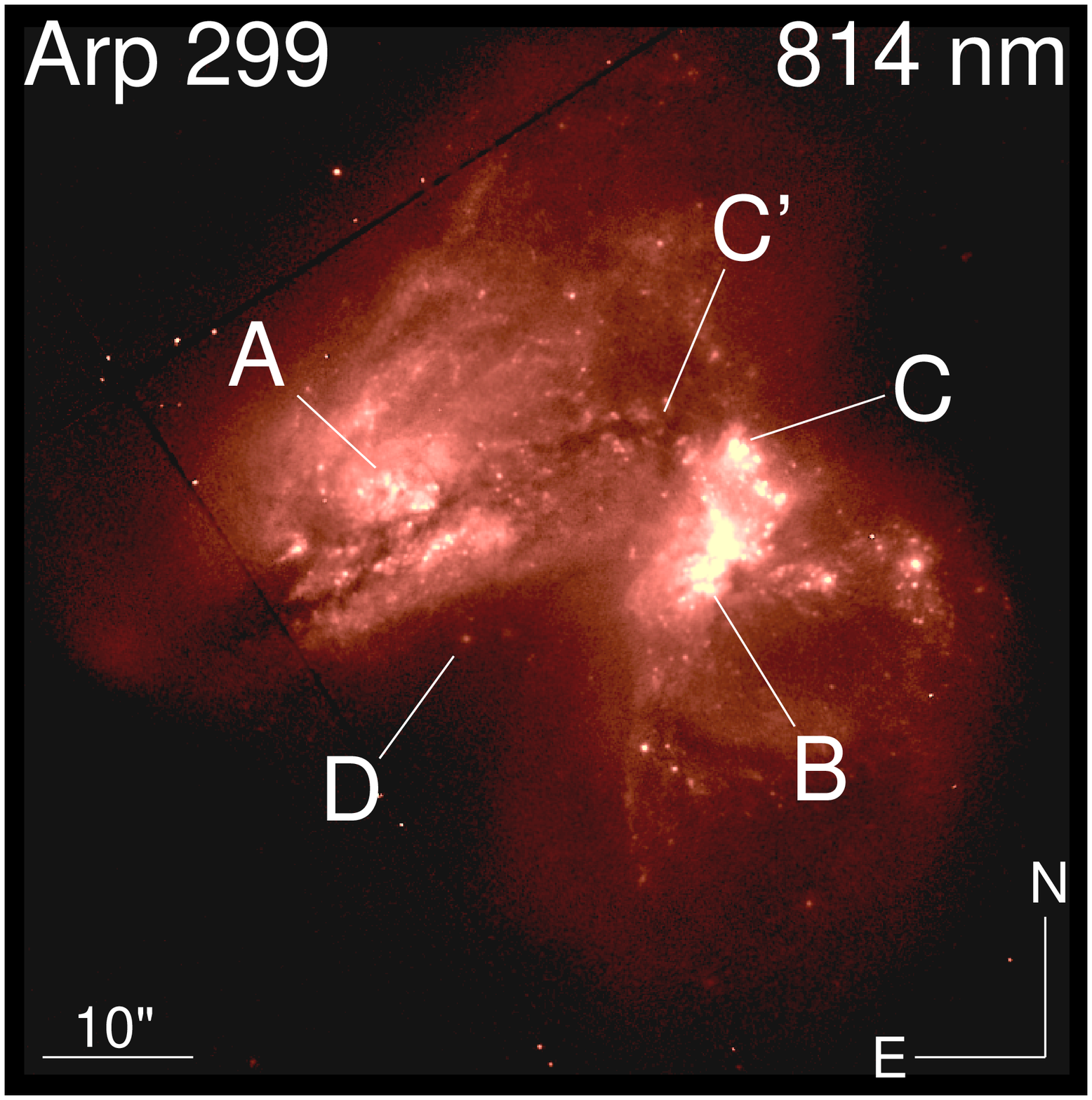}
\caption{
HST WFPC2 814 nm image of Arp~299, taken from \citet{nef04}. The 
primary merger nuclei A and B, the possible smaller nuclei C
and C$'$, and the additional radio source D are all labelled.
Reproduced by permission of the AAS.}
\label{fig:wfpc}
\end{figure}
\clearpage

\begin{figure}
\plotone{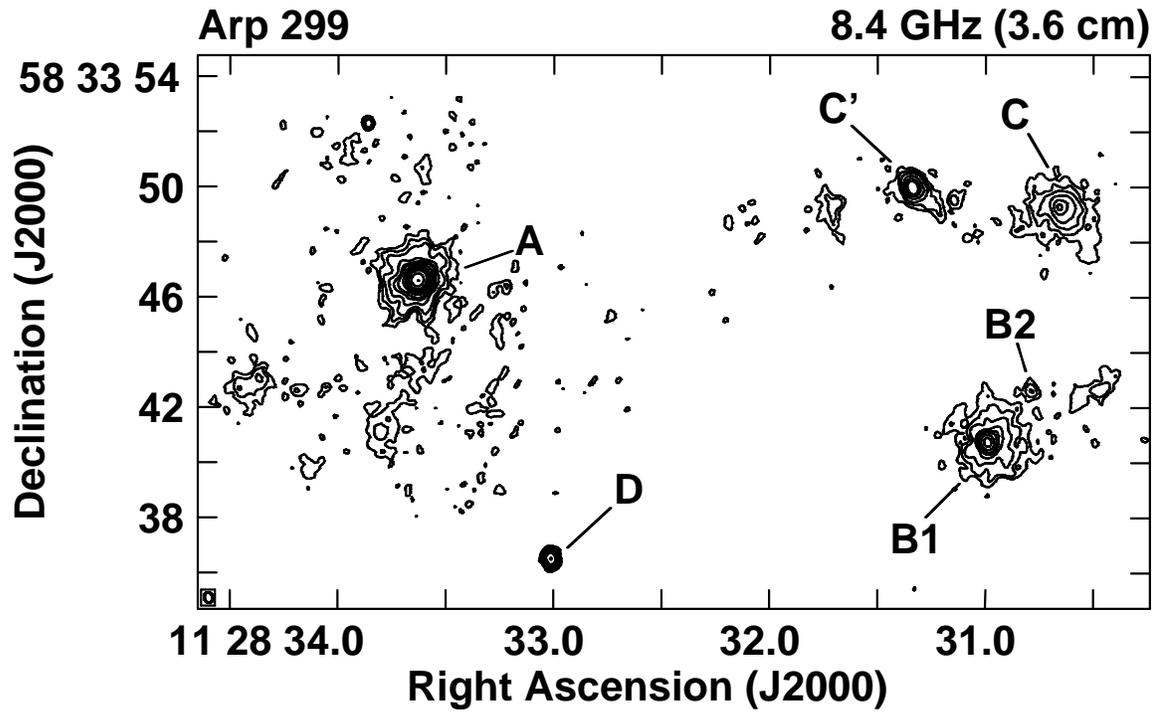}
\caption{
VLA 8.4~GHz image of Arp~299, taken from \citet{nef04}.
Labels are similar to those in Figure~1, except that nucleus B
is separated into sources B1 and B2. Reproduced by permission
of the AAS.}
\label{fig:vla-old}
\end{figure}
\clearpage

\begin{figure}
\plotone{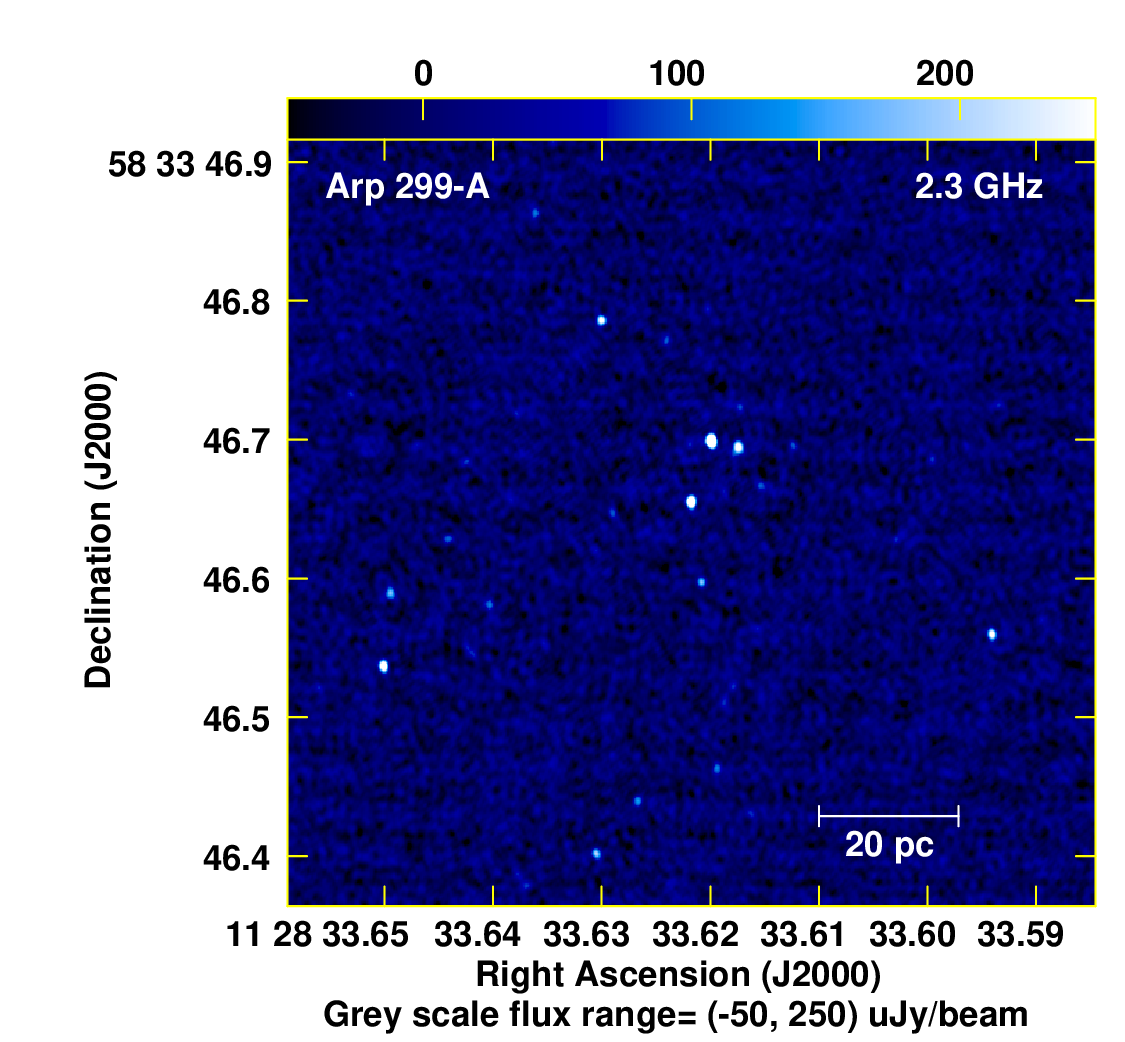}
\caption{
VLBI image of nucleus A of Arp 299 at 2.3 GHz, combining
data from all four epochs. The intensity scale is shown at the top.}
\label{fig:vlbiA-s}
\end{figure}
\clearpage

\begin{figure}
\plotone{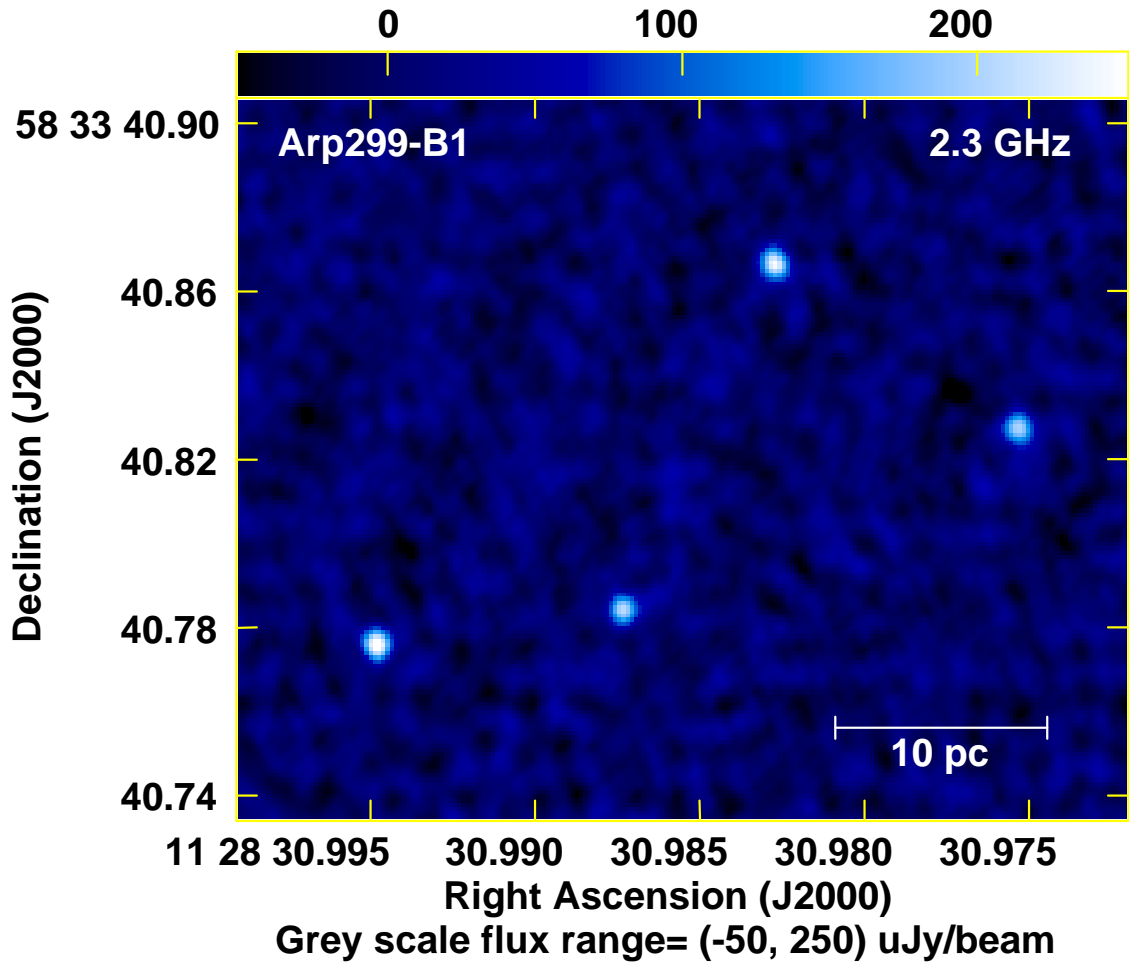}
\caption{
VLBI image of nucleus B1 of Arp 299 at 2.3 GHz, combining
data from all four epochs. The intensity scale is shown at the top.}
\label{fig:vlbiB1-s}
\end{figure}
\clearpage

\begin{figure}
\plotone{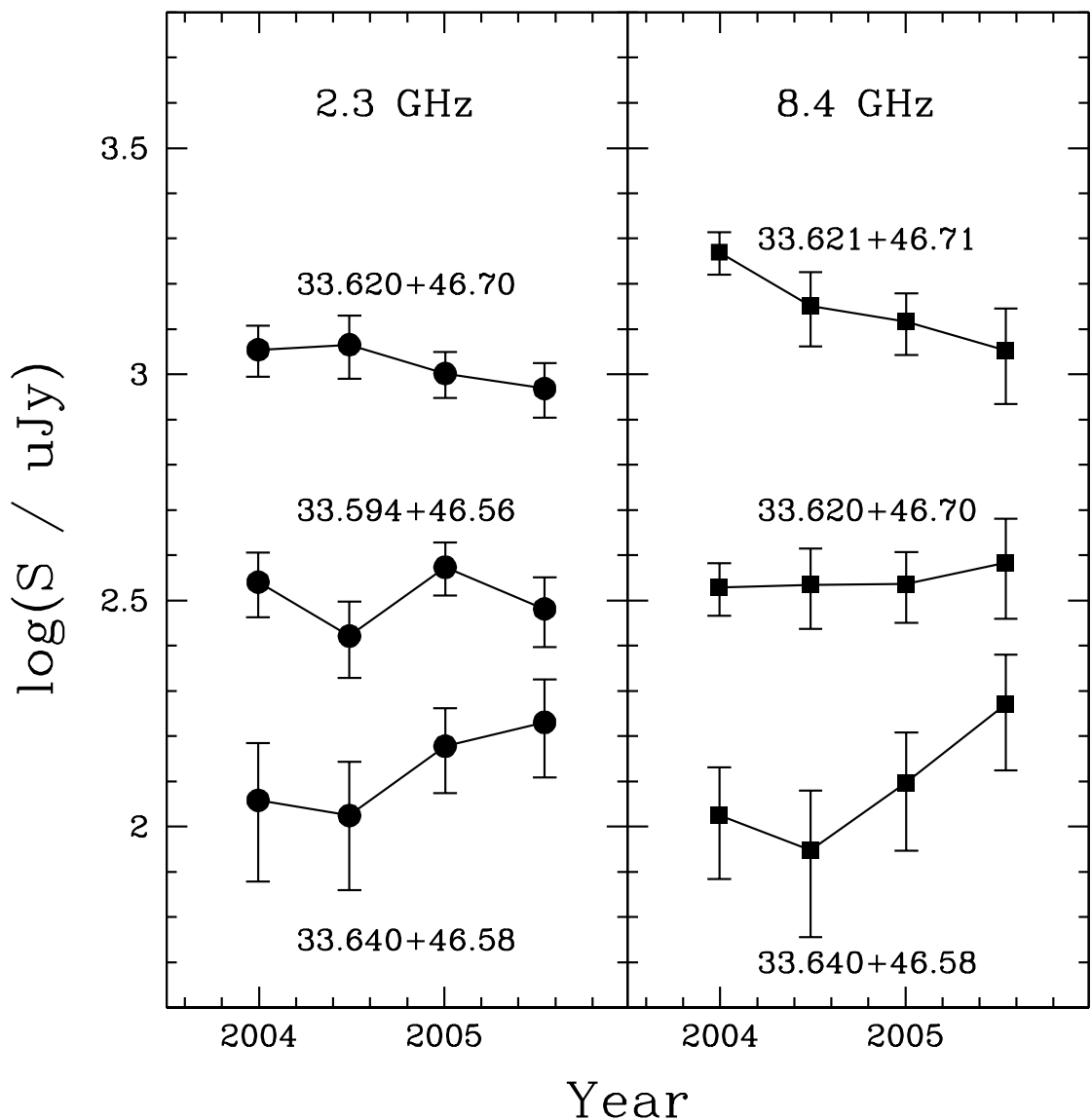}
\caption{
Time histories of strengths of compact sources at three
different flux-density levels, in nucleus A, at 2.3
GHz (left) and 8.4 GHz (right).  Each flux-density curve
is labelled by the source designation.  Two sources
(33.620+46.70 and 33.640+46.58) are shown in both panels.
The fading 8.4 GHz source, 33.621+46.71, is the apparent
supernova reported previously by \citet{nef04}.}
\label{fig:fluxtime}
\end{figure}
\clearpage

\begin{figure}
\plotone{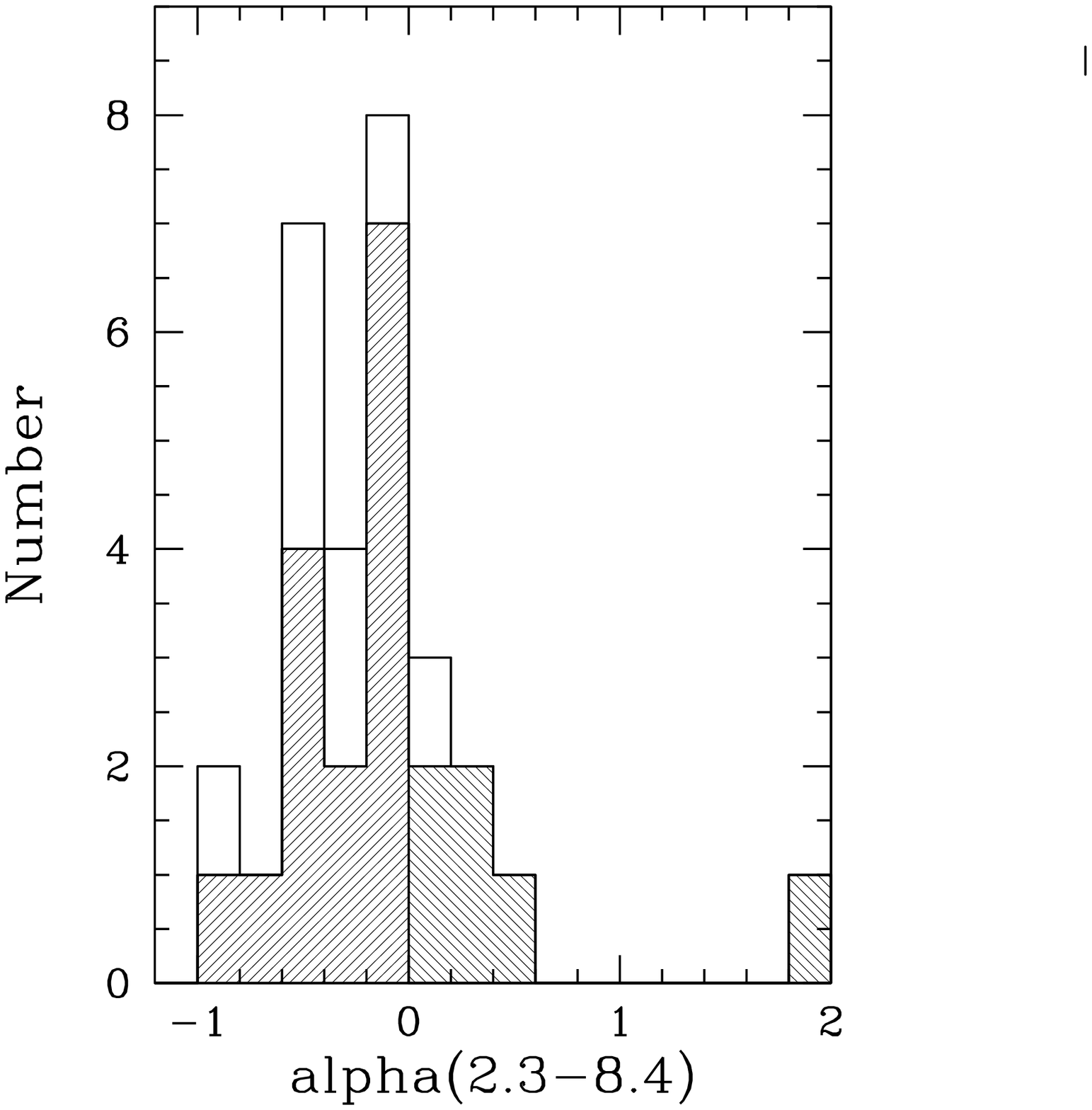}
\caption{
Distribution of two-point spectral indices of compact radio
sources in nucleus A. Hatched values represent either lower
limits ($\alpha > 0$) or upper limits ($\alpha < 0$), while
the eight sources detected at both frequency bands are
shown un-hatched.}
\label{fig:spectra}
\end{figure}
\clearpage

\begin{figure}
\plotone{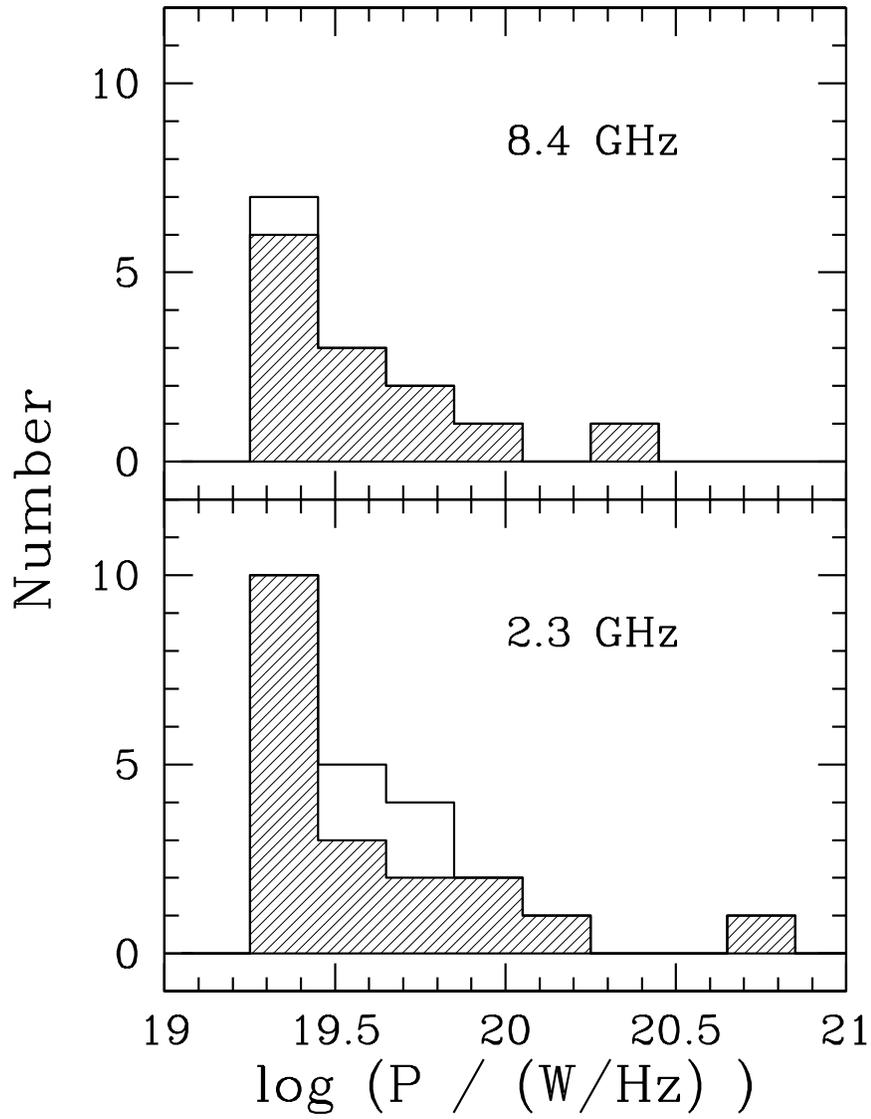}
\caption{
Histogram of radio luminosities of the compact radio sources
in the two nuclei of Arp~299.  The sources in nucleus A are
shaded, while those in nucleus B1 are unshaded. The recent 
supernova in nucleus B1 is not included.}
\label{fig:lumf}
\end{figure}
\clearpage

\begin{figure}
\plotone{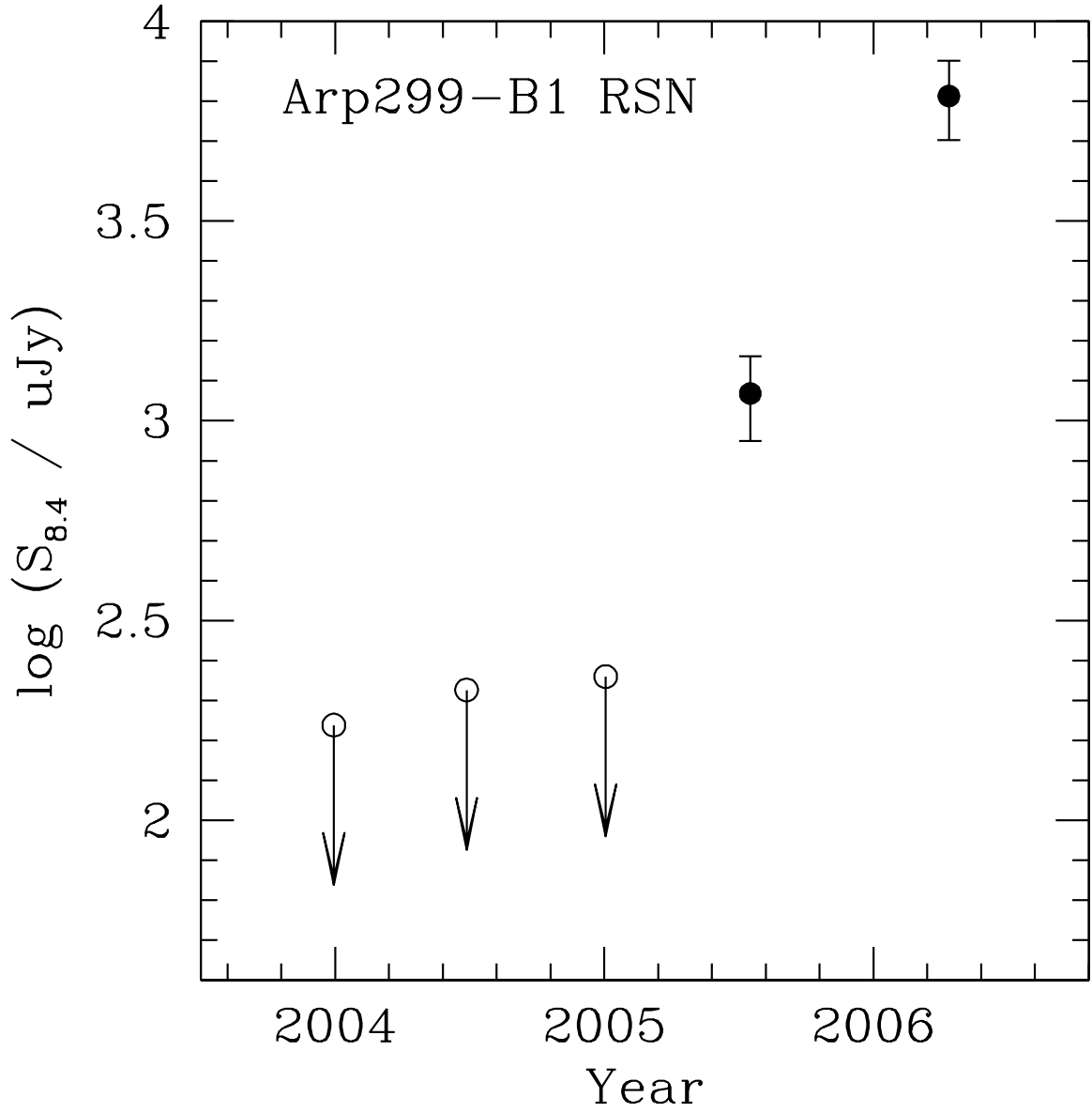}
\caption{
Light curve of radio supernova 30.988+40.78 
in nucleus B1 at 8.4 GHz, showing four
VLBI observations and the VLA archival observation from 2006.29.  Open
circles are $6\sigma$ upper limits from our VLBI observations, and
$1\sigma$ error bars are shown on the two detections.}
\label{fig:b1sn}
\end{figure}
%

\begin{deluxetable}{lcccc}
\tablecolumns{5}
\tablewidth{0pc}
\tablecaption{VLBA+GBT Observations of Arp~299}
\tablehead{
\colhead{(1)}& \colhead{(2)}& \colhead{(3)}& \colhead{(4)}&
\colhead{(5)} \\
\colhead{Date}&\colhead{Sky Frequency}&
\colhead{Integration}&\colhead{Resolution}&\colhead{rms noise\tablenotemark{a}} \\
\colhead{}&\colhead{(GHz)}&\colhead{(min)}&\colhead{(milliarcsec)}&
\colhead{($\mu$Jy beam$^{-1}$) } }
\startdata
2003 Dec 29&2.2715&178&6.80$\times$5.37&36 \\
&8.4215&180&1.93$\times$1.31&27\\
2004 Jun 27&2.2715&178&6.29$\times$4.69&29 \\
&8.4215&181&1.76$\times$1.14&27\\
2005 Jan 02&2.2715&172&6.37$\times$4.97&27 \\
&8.4215&173&2.16$\times$1.19&31\\
2005 Jul 17&2.2715&174&6.49$\times$4.85&35 \\
&8.4215&176&1.84$\times$1.16&30\\
\enddata
\label{tab:obs}
\tablenotetext{a}{The ``raw'' rms noise for each individual image is
given in this table.  These values should be scaled upward by the 
amplitude correction factors shown in Table~2 to estimate the ``true''
noise values.}
\end{deluxetable}

\begin{deluxetable}{lccccccc}
\tabletypesize{\scriptsize}
\tablecolumns{8}
\tablewidth{0pc}
\tablecaption{Amplitude Correction Factors Caused by Coherence Losses}
\tablehead{
\colhead{(1)}& \colhead{(2)}& \colhead{(3)}& \colhead{(4)}&
\colhead{(5)}& \colhead{(6)} &\colhead{(7)}& \colhead{(8)} \\
\colhead{Frequency}&\colhead{2003 Dec 29}&\colhead{2004 Jun 27}&
\colhead{2005 Jan 02}&\colhead{2005 Jul 17}&\colhead{Average}&
\colhead{Measured rms}&\colhead{Effective rms} \\
\colhead{}& \colhead{}& \colhead{}& \colhead{}&
\colhead{}& \colhead{}& \colhead{($\mu$Jy beam$^{-1}$)}& 
\colhead{($\mu$Jy beam$^{-1}$)} }
\startdata
2.3 GHz&1.15&1.24&1.11&1.19&1.17&18&21 \\
8.4 GHz&1.07&1.31&1.23&1.43&1.26&15&19 \\
\enddata
\label{tab:cohere}
\end{deluxetable}
\clearpage

\begin{deluxetable}{lcccl}
\tablecolumns{5}
\tablewidth{0pc}
\tablecaption{Compact Sources Detected at 2.3 GHz}
\tablehead{
\colhead{Name}&\colhead{$\alpha$ (J2000)} & \colhead{$\delta$ (J2000)} & \colhead{Flux Density} &
\colhead{Comments} \\
\colhead{}&\colhead{$11^{\rm h}28^{\rm m}$} & \colhead{58\arcdeg 33$'$} &
\colhead{($\mu$Jy)}&\colhead{} }
\startdata
\underbar{Nucleus A}& \\
33.594+46.56&33.5941$^{\rm s}$&46\farcs560&354$\pm$32&Also at 8.4 GHz\\
33.612+46.70&33.6124$^{\rm s}$&46\farcs696&119$\pm$22&\\
33.615+46.67&33.6154$^{\rm s}$&46\farcs667&125$\pm$23&\\
33.617+46.72&33.6173$^{\rm s}$&46\farcs724&115$\pm$22&\\
33.618+46.70&33.6175$^{\rm s}$&46\farcs695&382$\pm$33&\\
33.619+46.46&33.6194$^{\rm s}$&46\farcs463&149$\pm$23&\\
33.620+46.70&33.6200$^{\rm s}$&46\farcs699&1090$\pm$76&Also at 8.4 GHz\\
33.621+46.60&33.6209$^{\rm s}$&46\farcs597&217$\pm$26&\\
33.622+46.66&33.6218$^{\rm s}$&46\farcs655&662$\pm$49&Also at 8.4 GHz\\
33.624+46.77&33.6241$^{\rm s}$&46\farcs771&121$\pm$23&\\
33.627+46.44&33.6267$^{\rm s}$&46\farcs440&165$\pm$24&\\
33.629+46.65&33.6290$^{\rm s}$&46\farcs647&114$\pm$22&\\
33.630+46.79&33.6301$^{\rm s}$&46\farcs786&335$\pm$31&Also at 8.4 GHz\\
33.631+46.40&33.6305$^{\rm s}$&46\farcs402&241$\pm$27&\\
33.636+46.86&33.6361$^{\rm s}$&46\farcs863&139$\pm$23&\\
33.640+46.58&33.6404$^{\rm s}$&46\farcs581&141$\pm$23&Also at 8.4 GHz\\
33.644+46.63&33.6442$^{\rm s}$&46\farcs628&126$\pm$23&\\
33.650+46.59&33.6495$^{\rm s}$&46\farcs589&211$\pm$25&Also at 8.4 GHz\\
33.650+46.54&33.6501$^{\rm s}$&46\farcs537&487$\pm$39&Also at 8.4 GHz\\
\\
\underbar{Nucleus B1}& \\
30.975+40.83&30.9753$^{\rm s}$&40\farcs828&237$\pm$26&\\
30.983+40.87&30.9827$^{\rm s}$&40\farcs867&294$\pm$29&\\
30.987+40.78&30.9873$^{\rm s}$&40\farcs784&245$\pm$27&Near 8.4 GHz supernova\\
30.995+40.78&30.9948$^{\rm s}$&40\farcs776&332$\pm$31&Also at 8.4 GHz\\
\\
\underbar{Source D}& \\
33.011+36.55&33.0109$^{\rm s}$&36\farcs549&964$\pm$68&Background?\\
\enddata
\label{tab:sflux}
\end{deluxetable}
\clearpage
\begin{deluxetable}{lcccl}
\tablecolumns{5}
\tablewidth{0pc}
\tablecaption{Compact Sources Detected at 8.4 GHz}
\tablehead{
\colhead{Name}&\colhead{$\alpha$ (J2000)} & \colhead{$\delta$ (J2000)} & \colhead{Flux Density} &
\colhead{Comments} \\
\colhead{}&\colhead{$11^{\rm h}28^{\rm m}$} & \colhead{58\arcdeg 33$'$} &
\colhead{($\mu$Jy)}&\colhead{} }
\startdata
\underbar{Nucleus A}& \\
33.594+46.56&33.5941$^{\rm s}$&46\farcs560&194$\pm$26&Also at 2.3 GHz\\
33.599+46.64&33.5992$^{\rm s}$&46\farcs637&132$\pm$22&\\
33.620+46.70&33.6199$^{\rm s}$&46\farcs699&371$\pm$38&Also at 2.3 GHz\\
33.621+46.71&33.6212$^{\rm s}$&46\farcs707&1410$\pm$126&Declining flux\tablenotemark{a}\\
33.622+46.66&33.6218$^{\rm s}$&46\farcs655&504$\pm$49&Also at 2.3 GHz\\
33.628+46.62&33.6280$^{\rm s}$&46\farcs623&147$\pm$23&\\
33.630+46.79&33.6301$^{\rm s}$&46\farcs786&366$\pm$38&Also at 2.3 GHz\\
33.631+46.62&33.6307$^{\rm s}$&46\farcs620&214$\pm$27\\
33.636+46.68&33.6356$^{\rm s}$&46\farcs677&141$\pm$23&\\
33.639+46.55&33.6391$^{\rm s}$&46\farcs550&129$\pm$22&\\
33.640+46.58&33.6402$^{\rm s}$&46\farcs580&118$\pm$22&Also at 2.3 GHz\\
33.650+46.59&33.6495$^{\rm s}$&46\farcs591&159$\pm$24&Also at 2.3 GHz\\
33.650+46.54&33.6501$^{\rm s}$&46\farcs537&267$\pm$30&Also at 2.3 GHz\\
\\
\underbar{Nucleus B1}& \\
30.988+40.78&30.9875$^{\rm s}$&40\farcs784&1170$\pm$280&Fourth epoch only\\
30.995+40.78&30.9948$^{\rm s}$&40\farcs776&153$\pm$23&Also at 2.3 GHz\\
\\
\underbar{Source D}& \\
33.011+36.55&33.0108$^{\rm s}$&36\farcs549&1810$\pm$161&Background?\\
\enddata
\label{tab:xflux}
\tablenotetext{a}{The flux densities of source 33.621+46.71 at the four new
observing epochs were as follows. 2003 Dec 29: $1861\pm 199$~$\mu$Jy;
2004 Jun 27: $1416\pm 264$~$\mu$Jy; 2005 Jan 02: $1308\pm 204$~$\mu$Jy;
2005 Jul 17: $1128\pm 269$~$\mu$Jy. These also are plotted in the
upper right portion of Fig.~5.}
\end{deluxetable}

\begin{deluxetable}{lcc}
\tablecolumns{3}
\tablewidth{0pc}
\tablecaption{VLA 8.4-GHz Flux Densities of Arp 299}
\tablehead{
\colhead{Epoch}&\multicolumn{2}{c}{$S_{8.4}$ (mJy)} \\
\cline{2-3} \\
\colhead{}&\colhead{Nucleus A}&\colhead{Nucleus B1} }
\startdata
2004.84&$72.7\pm 3.6$&$7.1\pm 0.4$ \\
2006.29&$64.4\pm 6.4$&$13.6\pm 1.4$ \\

\enddata
\label{tab:vla}
\end{deluxetable}

\end{document}